\documentclass[12pt]{iopart}
\expandafter\let\csname equation*\endcsname\relax
\expandafter\let\csname endequation*\endcsname\relax
\usepackage{amsmath}
\usepackage{iopams}
\usepackage{xcolor}
\usepackage{graphicx}
\usepackage{cite}

\usepackage{mathtools}
\usepackage{physics}

\newtheorem{prob}{Problem}
\newtheorem{const}{Constraint}

\usepackage[colorlinks=true]{hyperref}
\usepackage{cleveref}
\begin{document}

\title[Biorthogonal Renormalization]{Biorthogonal Renormalization}

\author{Elisabet Edvardsson}
\address{Department of Physics, Stockholm University, AlbaNova University Center, 106 91 Stockholm, Sweden}

\author{J Lukas K König}
\address{Department of Physics, Stockholm University, AlbaNova University Center, 106 91 Stockholm, Sweden}

\ead{lukas.konig@fysik.su.se}

\author{Marcus St\aa{}lhammar}
\address{Nordita, KTH Royal Institute of Technology and Stockholm University, Hannes Alfv\'ens v\"ag 12, SE-106 91 Stockholm, Sweden}

\newcommand{\Lukas}[1]{\textcolor{orange}{#1}}
\newcommand{\Elisabet}[1]{\textcolor{blue}{#1}}
\newcommand{\Marcus}[1]{\textcolor{purple}{#1}}

\newcommand{\inner}[2]{\ensuremath{\left(#1,#2\right)}}
\newcommand{\simpleset}[1]{\ensuremath{\left\{#1\right\}}}
\vspace{10pt}
\begin{indented}
\item[]\today
\end{indented}

\begin{abstract}
The biorthogonal formalism extends conventional quantum mechanics to the non-Hermitian realm.
It has, however, been pointed out that the biorthogonal inner product changes with the scaling of the eigenvectors, an ambiguity whose physical significance is still being debated. 
Here, we revisit this issue and argue that this choice of normalization is of physical importance. 
We illustrate in which settings quantities such as expectation values and transition probabilities depend on the scaling of eigenvectors, and in which settings the biorthogonal formalism remains unambiguous. 
To resolve the apparent scaling ambiguity, we introduce an inner product independent of the gauge choice of basis and show that its corresponding mathematical structure is consistent with quantum mechanics.
Using this formalism, we identify a deeper problem relating to the physicality of Hilbert space representations, which we illustrate using the position basis. 
\end{abstract}

%
%
%
%
%

\section{Introduction}

The foundations of quantum mechanics rely on the Hermiticity constraint, which ensures that all operators related to physical observables have real spectra.
However, recent years have marked a paradigm shift as the study of non-Hermitian Hamiltonians has intensified greatly, both from an experimental and a theoretical point of view~\cite{bergholtzExceptionalTopologyNonHermitian2021,ashidaNonHermitianPhysics2020}. 
These operators serve as effective descriptions of systems subject to, e.g., dissipation or gain and loss, and are fundamentally different from their Hermitian counterparts, partly because they have complex spectra and different sets of left and right eigenvectors. 
Consequently, non-Hermitian operators display many physical features that have no Hermitian counterparts; arguably the most prominent and well-studied have been the breakdown of the bulk-boundary correspondence~\cite{yaoEdgeStatesTopological2018,kunstBiorthogonalBulkBoundaryCorrespondence2018a,edvardssonNonHermitianExtensionsHigherorder2019,ghatakObservationNonHermitianTopology2020} and the appearance of deficiencies at which both eigenvalues and eigenvectors coalesce, the so-called exceptional points~\cite{carlstromKnottedNonHermitianMetals2019,stalhammarHyperbolicNodalBand2019,stalhammarClassificationExceptionalNodal2021,mandalSymmetryHigherOrderExceptional2021a,sayyadRealizingExceptionalPoints2022a,sayyadSymmetryprotectedExceptionalNodal2022,delplaceSymmetryProtectedMultifoldExceptional2021}. 
Additional examples include the extended 38-fold symmetry classification~\cite{kawabataClassificationExceptionalPoints2019,gongTopologicalPhasesNonHermitian2018,zhouPeriodicTableTopological2019}, and the physical consequences of the respective symmetries \cite{wojcikHomotopyCharacterizationNonHermitian2020,liHomotopicalCharacterizationNonHermitian2021,huKnotsNonHermitianBloch2021a,huKnotTopologyExceptional2022,wojcikEigenvalueTopologyNonHermitian2022a}, where parity-time symmetry comprise one well studied case. While parity-time-symmetric operators can replace their Hermitian counterparts in an equivalent formulation of quantum mechanics, due to their capacity of hosting real eigenvalue spectra~\cite{benderRealSpectraNonHermitian1998,benderPTSymmetricQuantumMechanics1999}, they are today understood as effective descriptions of optical systems where the symmetry reflects a balance between gain and loss~\cite{ozdemirParityTimeSymmetry2019}.

The relaxation of the Hermiticity constraint has fundamental consequences on the underlying mathematical framework of the theory.
As an example, the previously mentioned different sets of left and right eigenvectors, $\{\ket{L_n}\}$ and $\{\ket{R_n}\}$, respectively, are no longer individually orthogonal. Instead, they are \emph{biorthogonal}, i.e., $\braket{L_n}{R_m}\propto\delta_{nm}$, and the notion of inner product has to be modified in order to make connections to, e.g., probability and projections. To fulfill this purpose, the non-Hermitian community is mainly employing what is called the \emph{biorthogonal inner product}~\cite{brodyBiorthogonalQuantumMechanics2014}. 
This inner product has several benefits and is of physical relevance as it may predict the (dis)appearance of boundary states in lattice models and can thus be used to formulate the biorthogonal bulk-boundary correspondence for non-Hermitian systems~\cite{kunstBiorthogonalBulkBoundaryCorrespondence2018a,edvardssonNonHermitianExtensionsHigherorder2019,edvardssonPhaseTransitionsGeneralized2020}.

Despite its range of successful applications, there is an ambiguity in defining the biorthogonal inner product, see, e.g., Ref.~\cite{ibanezAdiabaticityConditionNonHermitian2014a}. The normalization condition used in the theory developed in Ref.~\cite{brodyBiorthogonalQuantumMechanics2014}, henceforth referred to as the {\em biorthogonal formalism}, leaves a degree of freedom in how to pick the eigenvectors of the Hamiltonian; if $\ket{R_n}$ is multiplied by some number $c_n\in\mathbb{C}$, $\ket{L_n}$ can simply be rescaled by $1/c^*_n$ and still satisfy the normalization condition. 
Such a change of basis alters the biorthogonal inner product, making it apparent that its definition depends on the choice of eigenbasis. In situations where a single Hamiltonian is considered, this is not a problem as the choice of scaling of the eigenvectors merely determines how the states and obsrvables should be represented~\cite{brodyConsistencyPTsymmetricQuantum2016a}, but it becomes problematic when studying different Hamiltonians and comparisons between results are desired, for example through a shared position representation.
Examples include previous works in the biorthogonal bulk-boundary correspondence~\cite{kunstBiorthogonalBulkBoundaryCorrespondence2018a,edvardssonNonHermitianExtensionsHigherorder2019,edvardssonPhaseTransitionsGeneralized2020}, where the expectation value of the operator $\dyad{e_n}$, with $\ket{e_n}$ denoting the vector represented by $\begin{pmatrix} 0&\dots&0&1&0&\dots&0\end{pmatrix}^T$, is shown to be of physical importance despite its actual meaning being affected by how states are represented. This means that one needs to be careful in these and similar situations, as it is important that the physical meaning of the quantities remains the same when making comparisons between systems with different Hamiltonians. Similar problems are expected when studying the position representation of wave functions in the continuum case, as is done in, e.g., Ref.~\cite{silbersteinBerryConnectionInduced2020}, where the Berry connection is computed. It is further argued in Ref.~\cite{silbersteinBerryConnectionInduced2020} that when computing expectation values of position and momentum operators, the conventional Hermitian definition of expectation values is preferable over the notion stemming from the biorthogonal formalism.

To address the problems above, we revisit the biorthogonal formalism of non-Hermitian quantum mechanics in this work. 
We outline the potential problems and ambiguities of the formalism, with a particular focus on the biorthogonal inner product, and explain when they are of physical relevance. 
To eliminate these problems, we formulate a more general inner product in terms of an inner product matrix $G$ -- similar to what is done in Ref.~\cite{juNonHermitianHamiltoniansNogo2019} -- which is independent of the gauge choice of basis of the eigenvectors, leaving quantities such as the expectation value invariant under physically irrelevant choices.
While Ref.~\cite{juNonHermitianHamiltoniansNogo2019} discusses general matrices, we instead focus on specifying $G$ and argue that one particular such choice is favorable. 
Our work is of relevance in non-Hermitian physics as it explicitly suggests a simple, basis independent extension of the biorthogonal formalism, compatible with conventional quantum mechanics, that can directly be applied to physical setups. 
The topic of uniqueness has also been explored in the context of a metric operator formulation in Ref. \cite{scholtzQuasiHermitianOperatorsQuantum1992}.

The outline of the article is as follows. We start in Sec.~\ref{sec:background} by giving a short introduction to the biorthogonal formalism, discuss apparent problems with the formalism and when they are of physical importance. 
We especially focus on the biorthogonal inner product and its dependence on a gauge choice of basis vectors, which directly comprises the motivation for our work. 
In Sec.~\ref{sec:biip} we set out to resolve these problems. 
In particular, we define a new, basis independent, inner product and investigate its physical and mathematical properties. 
We sort out which problems in the biorthogonal formalism can be resolved in this way, and which remain. Sec.~\ref{sec:discussion} is devoted to a discussion about the result. We conclude in Sec.~\ref{sec:conclusion} by summarizing our most important results and suggesting future research directions within the field.

\section{Background and Motivation} \label{sec:background}
We start by introducing important concepts and present the main motivations for our work. In Sec.~\ref{sec:boip}, we give a brief survey the biorthogonal formalism outlined in Ref.~\cite{brodyBiorthogonalQuantumMechanics2014}. 
In Sec.~\ref{sec:compham}, we identify potential issues with the definition of the corresponding inner product and point out what problems may arise from it, while we show in Sec.~\ref{sec:expvales} why these issues have not caused problems in previous studies. 

The operators studied in this work are assumed to be linear, finite dimensional and free from eigenvalue degeneracies; in particular they are non-defective. 
The notations and other conventions set in this section will be used throughout the rest of the work, unless otherwise specified.

\subsection{The Biorthogonal Inner Product} \label{sec:boip}
Take as starting point two sets of vectors, denoted $\left\{\ket{R_n}\right\}$ and $\left\{\ket{L_n}\right\}$, that both span $\mathbb{C}^N$, but that do not necessarily consist of orthogonal vectors. Assume further that these two sets are biorthogonal to one another and that the vectors are scaled according to
\begin{equation} 
    \braket{L_m}{R_n} = \delta_{mn}.
    \label{eq:overlap-condition-LR}
\end{equation}
These sets can be used to introduce an inner product in the following way. For each vector 
\begin{equation}\label{eq:alphadef}
    \ket{\alpha} = \sum_n a_n\ket{R_n},
\end{equation}
define an \emph{associated vector},
\begin{equation}\label{eq:associated_state}
    \ket{\tilde{\alpha}} = \sum_n a_n\ket{L_n}.
\end{equation}
Then the \emph{biorthogonal inner product}, denoted by $(\cdot,\cdot)_B$, is defined in the following way:
\begin{equation} \label{eq:alphabeta}
    \left(\ket{\alpha},\ket{\beta}\right)_B = \braket{\tilde{\alpha}}{\beta}.
\end{equation}
Assuming that $\ket{\beta} = \sum_n b_n\ket{R_n}$, Eq.~\eqref{eq:alphabeta} becomes
\begin{equation}\label{eq:firstip}
    \left(\ket{\alpha},\ket{\beta}\right)_B = \sum_na_n^*b_n.
\end{equation}
As is argued in Ref.~\cite{brodyBiorthogonalQuantumMechanics2014}, this is a valid, positive definite, inner product and it is of particular use in systems described by non-Hermitian Hamiltonians; the left and right eigenvectors of such a Hamiltonian form two biorthogonal sets, and can thus be used to define a biorthogonal inner product. The vector space $\mathbb{C}^N$, in which the vectors $\ket{\alpha}$ and $\ket{\beta}$ live, together with the biorthogonal inner product, forms a Hilbert space, which is denoted by $\mathcal{H}_B$. Physical states will be represented by vectors in $\mathcal{H}_B$ and observables by operators acting on the space. 

As in the Hermitian case, the biorthogonal inner product can be used to compute probabilities, and the transition probability between states represented by $\ket{\alpha}$ and $\ket{\beta}$ is given by
\begin{equation}
    p_{\alpha\rightarrow\beta} = \frac{\braket{\tilde{\alpha}}{\beta}\braket{\tilde{\beta}}{\alpha}}{\braket{\tilde{\alpha}}{\alpha}\braket{\tilde{\beta}}{\beta}}.
\end{equation}
This is a number between $0$ and $1$ and, as argued in Refs.~\cite{brodyBiorthogonalQuantumMechanics2014,brodyConsistencyPTsymmetricQuantum2016a}, any choice of biorthogonal basis can be used to derive a consistent probability theory. Since the notion of probability exists, it is possible to also define expectation values. In the biorthogonal framework, the expectation value of an operator $Q$ in a state represented by the vector $\ket{\alpha}$ is introduced as,
\begin{equation} \label{eq:EVBO}
    \langle Q\rangle =\frac{\left(\ket{\alpha},Q\ket{\alpha}\right)_B}{\left(\ket{\alpha},\ket{\alpha}\right)_B} =  \frac{\mel{\tilde{\alpha}}{Q}{\alpha}}{\braket{\tilde{\alpha}}{\alpha}}.
\end{equation}
Any operator $Q$ can be written in the form
\begin{equation}
    Q = \sum_{mn} q_{mn}\dyad{R_m}{L_n},
\end{equation}
with $q_{mn}\in \mathbb{C}$. The operators for which the numbers $q_{mn}$ form a Hermitian matrix are called \emph{biorthogonally Hermitian}. For these operators, the expectation value given by Eq.~\eqref{eq:EVBO} is always real, and hence these operators are taken to correspond to observables and vice versa.

\subsection{A Physically Relevant Ambiguity} \label{sec:compham}
The normalization condition in Eq.~\eqref{eq:overlap-condition-LR} leaves a degree of freedom in the choice of eigenvectors of a non-Hermitian Hamiltonian. 
If $\left\{\ket{R_n}\right\}$ and $\left\{\ket{L_n}\right\}$ denote right and left eigenvectors of a non-Hermitian Hamiltonian that satisfies Eq.~\eqref{eq:overlap-condition-LR}, any other sets on the form $\left\{c_n\ket{R_n}\right\}$, $\left\{(c_n^*)^{-1 }\ket{L_n}\right\}$, with $c_n\in \mathbb{C}$, will also satisfy Eq.~\eqref{eq:overlap-condition-LR}. Thus, the biorthogonal inner product can in principle be defined using any of these sets of eigenvectors, and still satisfy the biorthonormality condition Eq.~\eqref{eq:overlap-condition-LR}. 
As is stated in Ref.~\cite{brodyBiorthogonalQuantumMechanics2014,brodyConsistencyPTsymmetricQuantum2016a}, this is not a problem when considering a single closed system, since the physical state can be represented by a different vector, leading to the same predictions. When comparing different systems to each other this might, however, matter. For example, it is common to consider a family of Hamiltonians that depend on some parameter $\gamma$ describing a lattice model of size $N$. This situation is studied in a non-Hermitian context in e.g. Refs.~\cite{kunstBiorthogonalBulkBoundaryCorrespondence2018a,edvardssonPhaseTransitionsGeneralized2020,yaoEdgeStatesTopological2018,edvardssonNonHermitianExtensionsHigherorder2019}. The biorthogonal inner product, and thus how physical properties of the system are evaluated, will depend both on $\gamma$ and $N$, but also on the choice of eigenvectors of $H(\gamma,N)$. 

When studying lattice models, the vector $\ket{e_n}$ is typically taken to represent the $n$th site in the lattice, for all values of $\gamma$ and $N$. This means that a physical meaning is assigned to the vector $\ket{e_n}$ without reference to an inner product. The question is now how this fits with the biorthogonal formalism. To investigate this, it is natural to study the consequences of a rescaling of the eigenvectors of the Hamiltonian for a fixed vector $\ket{\alpha}$. 
Take the vectors $\ket{\alpha}$  and $\ket{\beta}$ as in Sec.~\ref{sec:boip},  consider the previous sets of eigenvectors $\left\{\ket{R_n}\right\}$ and $\left\{\ket{L_n}\right\}$, and then introduce another set of eigenvectors as $\left\{\ket{R_n'}\right\} = \left\{c_n\ket{R_n}\right\}$ and $\left\{(c_n^*)^{-1 }\ket{L_n}\right\}$.
As noted earlier, this transformation gives another basis that is normalized as $\braket{L_m'}{R_n'} = \delta_{mn}$, just like the previous basis for which $\braket{L_m}{R_n} = \delta_{mn}$.
The two states $\ket{\alpha}$ and $\ket{\beta}$ can be expressed in either basis, using the coefficients
\begin{equation}
    \ket{\alpha} = \sum_n a_n\ket{R_n} =\sum_n \frac{a_n}{c_n}\ket{R_n'}\,\,\,\,\textrm{and}\,\,\,\,\ket{\beta} =\sum_n b_n\ket{R_n}=\sum_n \frac{b_n}{c_n}\ket{R_n'}.
\end{equation}
The associated vector  \(\ket{\tilde{\alpha}'}\), corresponding to this new basis, is different from $\ket{\tilde{\alpha}}$ defined in Eq.~\eqref{eq:associated_state}, and reads
\begin{equation}
    \ket{\tilde{\alpha}'} = \sum_n\frac{a_n}{c_n}\ket{L_n'},
\end{equation}
which means that the inner product defined by the new basis, denoted by $(\cdot,\cdot)_{B'}$, is given by
\begin{equation} \label{eq:secondip}
    (\ket{\alpha},\ket{\beta})_{B'} 
    = \braket{\tilde{\alpha}'}{\beta} 
    = \sum_n\frac{a_n^*b_n}{|c_n|^2}.
\end{equation}
Comparing Eqs.~\eqref{eq:firstip} and \eqref{eq:secondip}, we conclude that in general
\begin{equation}
    (\ket{\alpha},\ket{\beta})_B\neq (\ket{\alpha},\ket{\beta})_{B'},
\end{equation}
i.e., the inner product between two vectors is not kept invariant under a rescaling of the eigenvectors. The probability of measuring the energy $E_n$ of a particle whose state is represented by the vector $\ket{\alpha}$ corresponds in the two different inner products to $p_n$ and $p_n'$ to,  
\begin{align}
    p_n 
    &= \frac{\braket{L_n}{\alpha}\braket{\tilde{\alpha}}{R_n}}{\braket{\tilde{\alpha}}{\alpha}\braket{L_n}{R_n}} 
    = \frac{\abs{a_n}^2}{\sum_m \abs{a_m}^2},
    \\
    p_n' 
    &= \frac{\braket{L_n'}{\alpha}\braket{\tilde{\alpha}'}{R_n'}}{\braket{\tilde{\alpha}'}{\alpha}\braket{L_n'}{R_n'}} 
    = \frac{\abs{a_n}^2\abs{c_n}^{-2}}{\sum_m\abs{a_m}^2\abs{c_m}^{-2}},
\end{align}
which means that the probability depends on the choice of $c_n$.
The same thing holds true for expectation values. The expectation values of the Hamiltonian in the state represented by the vector $\ket{\alpha}$ using the two different inner products read
\begin{align}
    \langle H\rangle &= \frac{\sum_n E_n\abs{a_n}^2}{\sum_n\abs{a_n}^2}, \qq{and}
    \\
\label{eq:second_expvalue}
    \langle H\rangle' &= \frac{\sum_n E_n  \abs{a_n}^2\abs{c_n^{-2}}}{\sum_n\abs{a_n}^2\abs{c_n^{-2}}}.
\end{align}
This means that in general the expectation values $\langle H\rangle$ and $\langle H\rangle'$ differ even for biorthogonally Hermitian operators. This clearly leads to interpretational challenges when turning, for example, to the lattice models described previously. 

\begin{figure}
    \centering
    \includegraphics[scale=0.15]{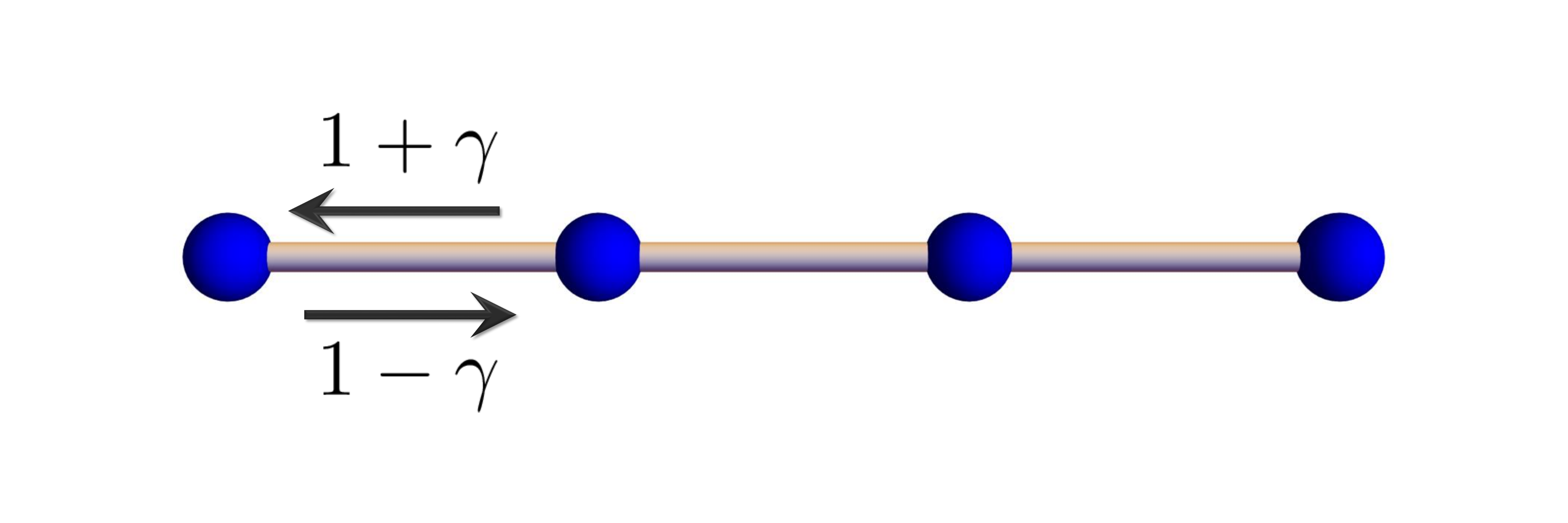}
    \caption{An illustration of the lattice model corresponding to the Hatano-Nelson Hamiltonian in Eq.~\eqref{eq:hatano_nelson}. As $\gamma\neq 0$, the right and left hopping amplitudes are different, resulting in a non-Hermitian Hamiltonian description.}
    \label{fig:hatano_nelson}
\end{figure}

\begin{figure}
    \centering
    \includegraphics[width=\textwidth]{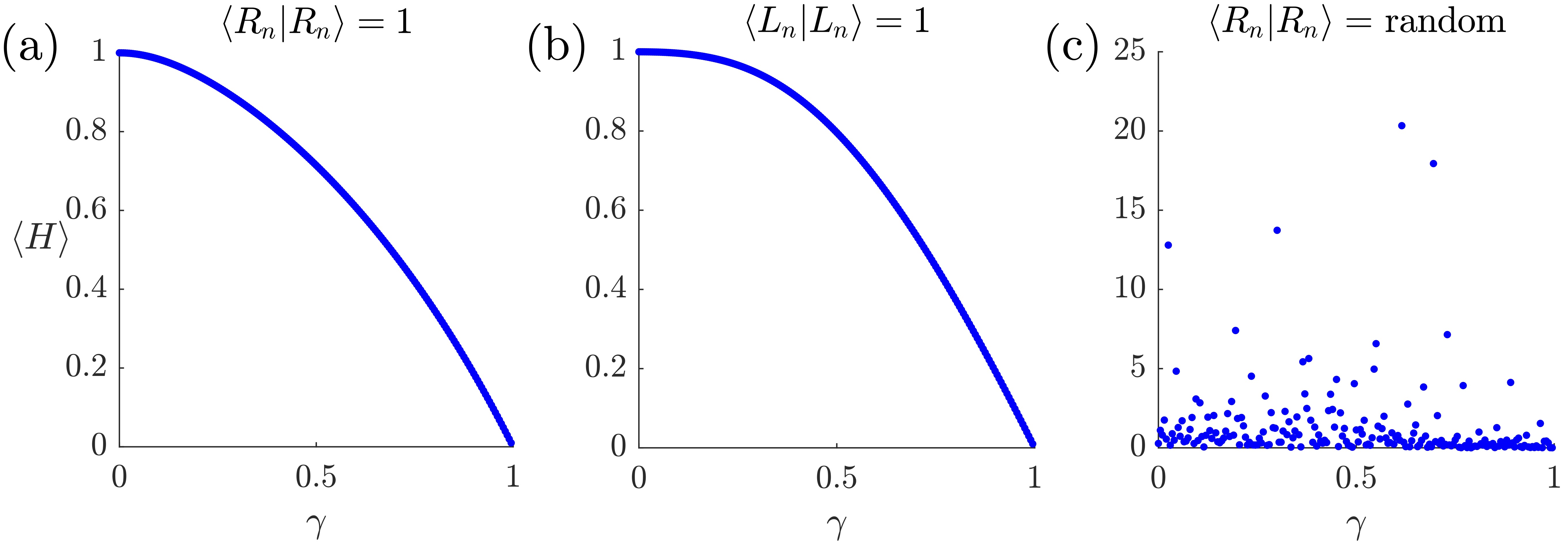}
    \caption{Absolute value of the expectation value of $H$ as a function of $\gamma$ in the state represented by $\ket{e_1}+\ket{e_2}$ using the biorthogonal inner product with three different normalization conditions. The inner product used in panel (a) is normalized according to $\braket{R_n}{R_n}=1$, in panel (b) according to $\braket{L_n}{L_n} = 1$, while in panel (c), the quantity $\braket{R_n}{R_n}$ takes random values. It should be noted that, despite being similar in shape, the graphs in panel (a) and (b) indeed differ from each other. The qualitative difference between the three panels indicates that the spare degrees of freedom caused by how eigenvectors can be rescaled in the biorthogonal formalism can affect physically relevant quantities.}
    \label{fig:expectation_values}
\end{figure}

We illustrate this using the concrete Hamiltonian
\begin{equation}\label{eq:hatano_nelson}
    H(\gamma) = \begin{pmatrix}
    0 & 1+\gamma &&\\
    1-\gamma & \ddots &\ddots &\\
    & \ddots &\ddots &1+\gamma\\
    && 1-\gamma &0
    \end{pmatrix},
\end{equation}
which describes the Hatano-Nelson model with open boundary conditions \cite{hatanoLocalizationTransitionsNonHermitian1996}. 
The lattice model associated with this Hamiltonian is shown in Fig.~\ref{fig:hatano_nelson}.
The left and right eigenvectors of $H(\gamma)$ can be used to define a biorthogonal inner product in several different ways according to the above reasoning. Here, we choose the following three options:
\begin{enumerate}
    \item Fix $\braket{R_n}{R_n} = 1$ and choose $\ket{L_n}$ such that $\braket{L_n}{R_n} = 1$.
    \item Fix $\braket{L_n}{L_n} = 1$ and choose $\ket{R_n}$ such that $\braket{L_n}{R_n} = 1$.
    \item For each $\gamma$ and $n$, fix $\braket{R_n}{R_n}$ to a random number between $0$ and $1$, and then pick $\ket{L_n}$ such that $\braket{L_n}{R_n} = 1$.
\end{enumerate}
The expectation values of $H(\gamma)$ in the state represented by $\ket{e_1}+\ket{e_2}$ in the different inner products are shown in Fig.~\ref{fig:expectation_values} as a function of $\gamma$. 
This shows that choosing different inner products can significantly change the qualitative shape of the curve, and that one has to be careful when making comparisons between different systems.

\subsection{Expectation Values in Eigenstates} \label{sec:expvales}
We have shown that the scaling of eigenvectors does affect physically relevant quantities like the energy expectation value when we fix a vector rather than a state. 
It is important to note, however, that if $\ket{\alpha}$ is an eigenvector of the Hamiltonian, the expectation value in the different inner products will be the same. Let $\ket{\alpha} = r\ket{R_n} = \frac{r}{c_n}\ket{R_n'}$ be an eigenvector of the Hamiltonian. Then
\begin{equation}
    \langle H\rangle' = \frac{E_n|r|^2|c_n|^{-2}}{|r^2||c_n|^{-2}} = \frac{|r|^2E_n}{|r^2|}= E_n = \langle H\rangle.
\end{equation}
This holds also for a general operator $Q$, as,
\begin{equation}
    \langle Q\rangle' = \frac{\mel{\tilde{\alpha}'}{Q}{\alpha}}{\braket{\tilde{\alpha}'}{\alpha}} = \frac{\mel{L_n}{Q}{R_n}|r|^2|c_n|^{-2}}{\braket{L_n}{R_n}|r|^2|c_n|^{-2}} = \langle Q\rangle.
\end{equation}
This is important, as it explains why most results in the literature are not affected by the ambiguity described in Sec.~\ref{sec:compham}. Examples include the expectation values of $\Pi_n=\dyad{e_n}$ computed in Refs.~\cite{edvardssonNonHermitianExtensionsHigherorder2019,kunstBiorthogonalBulkBoundaryCorrespondence2018a,edvardssonPhaseTransitionsGeneralized2020}, where they are used to formulate the biorthogonal bulk-boundary correspondence, and Berry connections~\cite{silbersteinBerryConnectionInduced2020,garrisonComplexGeometricalPhases1988,dattoliGeometricalPhaseCyclic1990}, which are both computed solely from the eigenstates of the respective systems. It is however important to stress that even though the expectation value of $\Pi_n$ in an eigenstate is unaffected by this problem, the physical meaning of its constituents, i.e., the vectors $\ket{e_n}$ is unclear and can depend on the choice of the corresponding eigenvectors. This becomes problematic when moving away from eigenstates, where the meaning of $\Pi_n$ depends on the choice of basis and the representation of vectors, such that its physical interpretation breaks down.

Although it is common to study properties of eigenvectors of the Hamiltonian in non-Hermitian systems, there is little work examining the superposition of eigenvectors. 
Such superpositions are central to a theory being quantum, and are regularly studied in the Hermitian case. 
The time evolution of a particle put on a specific lattice site is an example. 
Similar studies in non-Hermitian systems are desirable, but if these are to be carried out using the biorthogonal formalism, we have shown that problems may arise.
Instead, such future work requires a formalism that does allow for a consistent comparison of different systems. 

In summary, we have established that the representations of physical observables and states in the biorthogonal formalism depend on a choice of basis vectors and that this yields the following two problems:
\begin{prob} \label{prob:1}
Given a family of Hamiltonians $H(\gamma)$, the Hilbert space will change with $\gamma$ and physical states will thus be represented by different vectors depending on the choice of $\gamma$. Since the Hilbert space also depends on the choice of scaling of the eigenvectors of the Hamiltonian, comparing the physics of systems described by different Hamiltonians becomes difficult. 
\end{prob}
\begin{prob} \label{prob:2}
When studying non-Hermitian lattice models where Hamiltonians take the form of tight-binding matrices, the physical meaning of lattice position is implicitly assigned to the vectors $\ket{e_n}$. As the Hilbert space changes with the Hamiltonian and when choosing different eigenstates, the physical state that the vector $\ket{e_n}$ represents will also change. 
Thus the physical meaning of the vectors $\ket{e_n}$ will change even though they seem suitable for making physical predictions, e.g., predicting gap closings. 
\end{prob}
Recall that Problem~\ref{prob:2} is an instance of a larger class of problems related to any physical meaning that comes with the representation of a Hamiltonian.
In the context of real-eigenvalued Hamiltonians, this problem has been discussed using a metric operator formalism \cite{scholtzQuasiHermitianOperatorsQuantum1992}.
We now turn to how to solve these problems.

\section{Basis Independent Inner Product} \label{sec:biip}
In this section, we address Problems~\ref{prob:1} and \ref{prob:2}, and investigate if they can be solved by considering an inner product different from the one given by Eq.~\eqref{eq:alphabeta}. 
In Secs.~\ref{sec:indepIP}-\ref{sec:staterep} we consider Problem~\ref{prob:1}. 
We first introduce the inner product formalism that avoids the ambiguity discussed in the previous section (Sec.~\ref{sec:indepIP}).
We then study the notion of observables and the corresponding mathematical structure, and compare these to earlier interpretations (Sec.~\ref{sec:opalg}).
We go on to show that this inner product allows for representations of states to be mapped between different Hilbert spaces (Sec.~\ref{sec:staterep}).
Finally we treat Problem~\ref{prob:2} and the corresponding issues related to the physical meaning of position vectors $\ket{e_k}$ in Sec.~\ref{sec:physstates}.

\subsection{Reformulation} \label{sec:indepIP}
The scaling ambiguity related to Problem~\ref{prob:1} can be solved in several ways. One way would be to specify some eigenbasis that is to be used to construct associated states, say by requiring a fixed value of \(\braket{R_n}{R_n}\). Another way is to modify the inner product in such a way that it does not depend on the gauge choice of eigenbasis. Here we will pursue the latter. When constructing such an inner product, we impose the following constraints:
\begin{const} \label{const:1}
The inner product should admit a probabilistic interpretation.
\end{const}
\begin{const}\label{const:2}
The inner product should have the standard inner product as its Hermitian limit.
\end{const}
\begin{const}\label{const:3}
The inner product should be uniquely determined by the Hamiltonian.
\end{const}
Since a Hamiltonian admits several different inner products that satisfy the first two constraints, there can be multiple choices that also satisfy the third one. To choose between those, we require:
\begin{const}\label{const:4}
The inner product should be a natural choice.
\end{const}
To achieve the above, we find it beneficial to describe the inner product by a matrix $G$, such that it reads \(\inner{\ket{\alpha}}{\ket{\beta}}_G \coloneqq \mel{\alpha}{G}{\beta}\). Given a non-defective Hamiltonian, any set of left eigenvectors $\left\{\ket{L_n}\right\}$ is linearly independent and spans the vector space. Thus, the matrix $G$ can be expanded as,
\begin{equation}
    G = \sum_{mn} g_{mn}\dyad{L_m}{L_n}.
\end{equation}
Specifying the inner product corresponds to determining the constants $g_{mn}$. For the right eigenvectors of the Hamiltonian to be interpreted as stationary states, we require
\begin{equation}
    \mel{R_k}{G}{R_l} \propto \delta_{kl},
\end{equation}
for all $k,l$. This implies
\begin{equation}
    g_{kl}\braket{R_k}{L_k}\braket{L_l}{R_l} \propto \delta_{kl},
\end{equation}
which means that $g_{kl} = 0$ if $k\neq l$. Thus the matrix of the inner product takes the form
\begin{equation}\label{eq:inner_product_def}
    G = \sum_{n} g_{n}\dyad{L_n}{L_n}.
\end{equation}
This is similar to the form of the metric described in Refs.~\cite{brodyBiorthogonalQuantumMechanics2014, chenMathematicalFormalismNonHermitian2022}. To achieve a well-defined and positive definite inner product, $G$ must be Hermitian and all $g_n>0$, which additionally means it is invertible. Consequently, $G$ is in fact a Gram matrix. The particular case when all $g_n=1$ corresponds to the biorthogonal inner product in the biorthogonal formalism introduced in Ref.~\cite{brodyBiorthogonalQuantumMechanics2014}, which is not invariant under rescaling of the eigenvectors and thus failing to fulfill Constraint~\ref{const:3}.

Let us now address the points Constraints~\ref{const:1}-\ref{const:4} above, starting with Constraint~\ref{const:1}. 
A physical state \(\Psi\) is defined by its probability amplitudes that when squared give the probabilities for different measurement outcomes. Assuming the state $\Psi$ has probability amplitudes \(c_n\) corresponding to measurement of the energies $E_n$, the representation of $\Psi$ in the Hilbert space generated by the inner product defined by Eq.~\eqref{eq:inner_product_def}, becomes
\begin{equation}
	\ket{\Psi} = \sum_n c_n \frac{\ket{R_n}}{\sqrt{\inner{\ket{R_n}}{\ket{R_n}}_G}},
\end{equation}
where the fact that $\braket{L_m}{R_n}\propto\delta_{mn}$ ensures that $c_n$ can be computed from
\begin{equation}
    c_n = \frac{\inner{\ket{R_n}}{\ket{\Psi}}_G}{\sqrt{\inner{\ket{R_n}}{\ket{R_n}}_G}}.
\end{equation}
This means that, for any choice of constants \(g_n\), a state described by a set of probability amplitudes can be represented by a vector. Only the representation of this state depends on the choice of \(g_n\).

We emphasize that the inner product defined by Eq.~\eqref{eq:inner_product_def} keeps the notion of stationarity. As right eigenvectors are orthogonal in this inner product, $\inner{\ket{R_n}}{\ket{R_m}}_G\propto \delta_{mn}$, transition probabilities between two different eigenstates vanish.

Let us now address the remaining constraints. Constraint~\ref{const:3} reduces to that the inner product should be independent of the scale of $\ket{L_n}$ and $\ket{R_n}$ (which the biorthogonal formalism generally is not, as shown in Sec.~\ref{sec:background}). There are many inner products that satisfy this, so we now turn to Constraint~\ref{const:4} and choose
\begin{equation} \label{eq:finalgram}
	G = \sum_n \frac{\braket{R_n} }{\abs{\braket{L_n}{R_n}}^2}\ketbra{L_n}{L_n}.
\end{equation}
That this is a natural choice, comes from the fact that the Hamiltonian can be written in terms of projectors $P_n$, 
\begin{equation}
    H = \sum_n E_n P_n,
\end{equation}
where $P_n$ is an operator projecting onto the $n$th eigenstate. More precisely, these operators satisfy $P_mP_n = \delta_{mn}P_n$ and can be represented in terms of eigenstates of $H$ as $P_n = \ketbra{R_n}{L_n}/\braket{R_n}{L_n}$. In terms of projectors, the matrix $G$ can be written as
\begin{equation}
    G = \sum_n P_n^{\dagger}P_n.
\end{equation}
It is therefore a natural construction of a Hermitian positive definite matrix, given a Hamiltonian. 
The corresponding representation of a state with probability amplitudes $c_n$ is with this particular choice of $G$ given by
\begin{equation}\label{eq:representation}
    \ket{\Psi} = \sum_n c_n\frac{\ket{R_n}}{\sqrt{\braket{R_n}{R_n}}}.
\end{equation}

The inner product Eq.~\eqref{eq:finalgram} is independent of the scaling of eigenvectors and a natural choice given a specific Hamiltonian. Furthermore, in the Hermitian limit, the matrix of the inner product becomes the identity matrix, $G=\mathbf{1}$, which means that the inner product reduces to the standard one when the considered system is Hermitian. Thus, the choice of 
\begin{equation}
    g_n = \frac{\braket{R_n} }{\abs{\braket{L_n}{R_n}}^2}
\end{equation}
fulfills Constraints~\ref{const:1}-\ref{const:4}. This is the inner product which we consider for the remainder of this work.

\begin{figure}
    \centering
    \includegraphics[width=\linewidth]{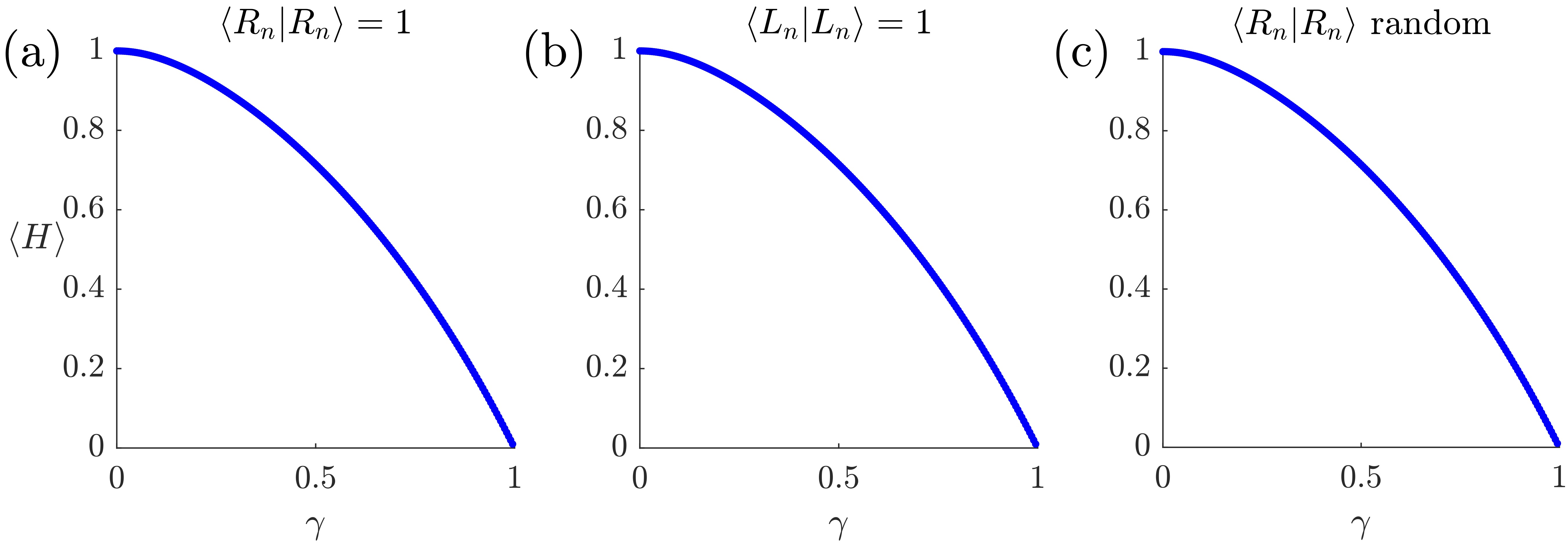}
    \caption{Expectation value in the inner product in Eq.~\eqref{eq:finalgram} of the Hamiltonian in the state represented by the vector $\ket{e_1}+\ket{e_2}$ as a function of $\gamma$ for three different choices of eigenvectors. Here, the expectation value does not change with the scaling of the eigenvectors, leaving an unambiguous interpretation of quantities computed from the inner product.}
    \label{fig:expvalues_with_G}
\end{figure}

To illustrate the benefit of using this inner product instead of the one used in the biorthogonal formalism, Fig.~\ref{fig:expvalues_with_G} shows the expectation value of the Hamiltonian in Eq.~\eqref{eq:hatano_nelson} in the state represented by the vector $\ket{e_1}+\ket{e_2}$, computed using the inner product defined in Eq.~\eqref{eq:finalgram}. Just a in Fig.~\ref{fig:expectation_values}, this is done for three different choices of eigenvectors, but contrary to what is seen in Fig~\ref{fig:expectation_values}, the expectation values displayed in Fig.~\ref{fig:expvalues_with_G} shows no dependence on the choice of normalization of the eigenvectors of the Hamiltonian.
Thus, this provides a consistent way of comparing results for different $\gamma$. 

Lastly, it is important to note that when only eigenstates are considered, the inner product defined in Eq.~\eqref{eq:inner_product_def} gives the same result as the biorthogonal inner product. 
As we have noted before, this is why the inner product ambiguity has not disrupted previous work.

\subsection{Observables and Self-Adjoint Operators} \label{sec:opalg}
So far, we have mainly been discussing the representation of states, but the representation of observables is equally important. In standard quantum mechanics, observables are represented by Hermitian operators, motivated by their real spectra. 
When considering the Hilbert space corresponding to the inner product defined by Eq.~\eqref{eq:finalgram}, this notion has to be appropriately changed. 
A natural generalization is comprised of operators that are self-adjoint with respect to this inner product, since such operators also have purely real eigenvalues. 
The adjoint \(Q^\star\) of some operator \(Q\) is defined by the action on arbitrary vectors via 
\begin{equation}\label{eq:addef}
    \inner{\ket{\alpha}}{Q^\star \ket{\beta}}_G:=\inner{Q\ket{\alpha}}{\ket{\beta}}_G,
\end{equation}
which means that the adjoint operator can be written as
\begin{equation}
    Q^\star = G^{-1} Q^\dagger G.
\end{equation}
The self-adjoint operators are thus operators satisfying
\begin{equation} \label{eq:saO}
Q=Q^\star = G^{-1} Q^\dagger G,
\end{equation}
with $G$ the matrix of the corresponding inner product. Thus, observables in the Hilbert space corresponding to the inner product $(\cdot,\cdot)_G$ are represented by operators satisfying Eq.~\eqref{eq:saO}. When considering the Hermitian limit, i.e., when $G$ converges to the identity operator, Eq.~\eqref{eq:saO} corresponds exactly to the notion of Hermitian operators with respect to the standard inner product. Thus, identifying observables in this way not only provides a notion from which the concepts of Hermitian quantum mechanics can be retrieved as a special case, but also constitutes a natural basis independent extension of the notion of biorthogonal Hermiticity introduced in Ref.~\cite{brodyBiorthogonalQuantumMechanics2014}.

Before we continue, we make a note that together with the operator norm induced by the inner product, the adjoint in Eq.~\eqref{eq:addef} defines a C\(\star\)-algebra which is the same structure that underlies traditional quantum mechanics~\cite{bratteliOperatorAlgebrasQuantum1979,bratteliOperatorAlgebrasQuantum1997}. Importantly, it provides the probability interpretation, as also explicitly shown in Sec.~\ref{sec:indepIP}, as well as the concepts of states and observables. 
States in this formalism arise as positive, linear functionals of unit norm, which for our purposes reduces to vectors normalized in the norm induced by the new inner product. The notion of observables is directly related to the definition of adjointness, as observables in a C\(\star\)-algebra correspond to operators that are self-adjoint.

\subsection{Mapping States and Observables Between Different Hilbert Spaces} \label{sec:staterep}

Suppose that a Hamiltonian can be represented by two operators $H_1$ and $H_2$ on the Hilbert spaces $\mathcal{H}_{G_1}$ and $\mathcal{H}_{G_2}$, respectively. Note that $H_1$ and $H_2$ share eigenvalues as they represent the same observable. Denote the respective eigenvectors by $\left\{\ket{R_n^{(1)}}\right\}$, $\left\{\ket{L_n^{(1)}}\right\}$ and $\left\{\ket{R_n^{(2)}}\right\}$, $\left\{\ket{L_n^{(2)}}\right\}$, normalized according to,
\begin{equation}
    \ket{R_n^{(1/2)}} = \frac{\ket{R_n^{(1/2)}}}{\sqrt{\braket{R_n^{(1/2)}}{R_n^{(1/2)}}}}, \quad \braket{L_n^{(1)}}{R_n^{(1)} } = \braket{L_n^{(2)}}{R_n^{(2)}}. 
\end{equation}
This gives rise to two different inner products, $G_1$ and $G_2$, and thus two different representations of states and observables. The representations have to be compatible for the theory to be consistent; there has to exist a map from $\mathcal{H}_{G_1}$ to $\mathcal{H}_{G_2}$ (and the other way around), preserving observables and states. Now we show that the map $\mathcal{T}:\mathcal{H}_{G_1}\rightarrow\mathcal{H}_{G_2}$, defined as,
\begin{equation} \label{eq:TRvec}
    \mathcal{T}\ket{R_n^{(1)}} = \ket{R_n^{(2)}},
\end{equation}
fulfills exactly that. 
As $\ket{R_n^{(1/2)}}$ spans $\mathcal{H}_{G_{1/2}}$, $\mathcal{T}$ is invertible and it is thus straight-forward to show,
\begin{equation} \label{eq:TLvec}
    \bra{L_n^{(1)}}\mathcal{T}^{-1} = \bra{L_n^{(2)}}, \quad G_1 = \mathcal{T}^{\dagger}G_2\mathcal{T}.
\end{equation}

Consider the transformation of states first. Suppose that there is a state represented by the vector $\ket{\alpha}\in\mathcal{H}_{G_1}$, defined as
\begin{equation}
    \ket{\alpha} = \sum_n c_n\ket{R_n^{(1)}}.
\end{equation}
If $E_n$ is the eigenvalue corresponding to $\ket{R_n^{(1)}}$, then the probability of measuring the energy $E_n$ is given by $|c_n|^2$ according to Eq.~\eqref{eq:representation}. $\mathcal{T}$ acts on $\ket{\alpha}$ as
\begin{equation}
    \mathcal{T}\ket{\alpha} = \sum_nc_n\ket{R_n^{(2)}},
\end{equation}
where $\mathcal{T}\ket{\alpha}\in\mathcal{H}_{G_2}$. Since $H_1$ and $H_2$ share eigenvalues, and $E_n$ thus is the eigenvalue corresponding also to $\ket{R_n^{(2)}}$, the probability of measuring the energy $E_n$ is still $|c_n|^2$, meaning that $\mathcal{T}$ preserves the probability notion, and hence the states.

Let us now turn to the case of operators. Let $Q$ be a representation of an observable on $\mathcal{H}_{G_1}$. By definition, $Q$ is self-adjoint in $\mathcal{H}_{G_1}$ with respect to the inner product $G_1$, meaning that
\begin{equation}
    G_1Q = Q^{\dagger}G_1.
\end{equation}
The operator $\mathcal{T}$ acts on $Q$ as $\mathcal{T}: Q\mapsto \mathcal{T}Q\mathcal{T}^{-1}$.
We show that $\mathcal{T}Q\mathcal{T}^{-1}$ is an observable on $\mathcal{H}_{G_2}$. 
Acting with $G_2$ from the left yields,
\begin{equation}
\begin{split}
    G_2\mathcal{T}Q\mathcal{T}^{-1} &= (\mathcal{T}^{\dagger})^{-1}G_1\mathcal{T}^{-1}\mathcal{T}Q\mathcal{T}^{-1} = (\mathcal{T}^{\dagger})^{-1}Q^{\dagger}G_1\mathcal{T}^{-1}\\
    &=(\mathcal{T}^{\dagger})^{-1}Q^{\dagger}\mathcal{T}^{\dagger}(\mathcal{T}^{\dagger})^{-1}G_1\mathcal{T}^{-1} = (\mathcal{T}Q\mathcal{T}^{-1})^{\dagger}G_2,
\end{split}
\end{equation}
which means that $\mathcal{T}Q\mathcal{T}^{-1}$ is self-adjoint in the inner product $G_2$ and thus represents an observable in $\mathcal{H}_{G_2}$. Importantly, $Q$ and $\mathcal{T}Q\mathcal{T}^{-1}$ represent the {\em same} observable in $\mathcal{H}_{G_1}$ and $\mathcal{H}_{G_2}$, respectively, since $Q$ and $\mathcal{T}Q\mathcal{T}^{-1}$ have the same eigenvalues and expectation values. 
The latter can be seen by making a direct calculation of the expectation value of $Q$ in the state represented by $\ket{\alpha}$ in $\mathcal{H}_{G_1}$, which gives
\begin{equation}
    \frac{\mel{\alpha}{G_1Q}{\alpha}}{\mel{\alpha}{G_1}{\alpha}} = \frac{\mel{\alpha}{\mathcal{T}^{\dagger}G_2\mathcal{T}Q\mathcal{T}^{-1}\mathcal{T}}{\alpha}}{\mel{\alpha}{\mathcal{T}^{\dagger}G_2\mathcal{T}}{\alpha}},
\end{equation}
where the right hand side corresponds exactly to the expectation value of $\mathcal{T}Q\mathcal{T}^{-1}$ in the state represented by $\mathcal{T}\ket{\alpha}$ in $\mathcal{H}_{G_2}$.

We can therefore conclude that under a transformation $\mathcal{T}$, the expected transformation of vectors and operators, i.e., 
\begin{equation}
    \mathcal{T}:Q\mapsto\mathcal{T}Q\mathcal{T}^{-1},\,\,\,\,\text{and}\,\,\,\,\,\ \mathcal{T}:\ket{\alpha}\mapsto\mathcal{T}\ket{\alpha}.
\end{equation}
preserves the notions of states and observables in different Hilbert spaces.
This also extends to representatives of the Hamiltonian; \(\mathcal{T}H_1\mathcal{T}^{-1} = H_2\), and hence the map \(\mathcal{T}\) also induces representations on different Hilbert spaces; the two notions are equivalent.

To conclude, the above shows that for inner products of the form of Eq.~\eqref{eq:finalgram}, there exist well-defined linear transformations that map the representations of both states and observables from one Hilbert space to another.
Consequently, this allows us to interpret the meaning of states and observables in various different Hilbert spaces.
In combination with the reasoning and results derived in Secs.~\ref{sec:indepIP} and \ref{sec:opalg}, this provides a solution to Problem~\ref{prob:1}.

\subsection{The Physical Meaning of Vectors} \label{sec:physstates}

As Secs.~\ref{sec:indepIP}-\ref{sec:staterep} have dealt with Problem~\ref{prob:1}, we now turn to Problem~\ref{prob:2} and the physical meaning of the vectors $\ket{e_k}$. Concretely, we want to answer the question of whether or not the {\em physical meaning} of vectors $\ket{e_k}$, i.e., the {\em physical state} that the vector $\ket{e_k}$ represents, can be transferred from one Hilbert space to another. It is clear that the inner product defined in Eq.~\eqref{eq:finalgram} does not preserve the meaning of these vectors, and the question is if it is possible to do it by choosing a different inner product. As previously argued, this question is of relevance in several physical setups, where quantities related to the expectation value of the projection operator $\Pi_k=\dyad{e_k}$ are computed, e.g., to predict gap closings and the existence of boundary states in lattice models.

To investigate this, recall that the norm and relation to other states and operators are central to the notion of a quantum state. 
It is therefore natural to check if the norm of $\ket{e_k}$ and the overlap between $\ket{e_k}$ and $\ket{e_l}$ can be preserved when the Hilbert space is changed. Consider the two Hamiltonians $H_1$ and $H_2$ of dimension $N$ and let them have eigenstates $\ket{R_n^{(1)}},\ket{L_n^{(1)}}$ and $\ket{R_n^{(2)}},\ket{L_n^{(2)}}$, respectively. Define the inner product matrices 
\begin{equation}
    G_1 = \sum_n g_n^{(1)}\dyad{L_n^{(1)}},\,\,\,\,\text{and}\,\,\,\,\,G_2 = \sum_n g_n^{(2)}\dyad{L_n^{(2)}},
\end{equation}
and let the vectors $\ket{e_k}$ be given by
\begin{equation} \label{eq:posvecaxp}
    \ket{e_k} = \sum_n c_{kn}^{(1)}\ket{R_n^{(1)}} = \sum_n c_{kn}^{(2)}\ket{R_n^{(2)}},
\end{equation}
Assume now that the set of $g_n^{(1)}$ is given. The question posed above then boils down to whether or not it is possible to choose the constants $g_n^{(2)}$ such that, for every $k,l \in \mathbb{Z}_+$,
\begin{align}
\mel{e_k}{G_1}{e_l} &= \mel{e_k}{G_2}{e_l}.
\end{align}
Inserting Eq.~\eqref{eq:posvecaxp} yields,
\begin{align}
    \sum_n g_n^{(1)}|c_{kn}^{(1)}\braket{L_n^{(1)}}{R_n^{(1)}}|^2 &= \sum_n g_n^{(2)}|c_{kn}^{(2)}\braket{L_n^{(2)}}{R_n^{(2)}}|^2, \quad l=k, \label{eq:normpres}
    \\
    \sum_n g_n^{(1)}\left(c_{kn}^{(1)}\right)^*c_{ln}^{(1)}|\braket{L_n^{(1)}}{R_n^{(1)}}|^2 &= \sum_n g_n^{(2)}\left(c_{kn}^{(2)}\right)^*c_{ln}^{(2)}|\braket{L_n^{(2)}}{R_n^{(2)}}|^2, \quad l\neq k, \label{eq:overlappres}
\end{align}
Eq.~\eqref{eq:normpres} corresponds to a system of equations whose size equals the dimension of the Hamiltonian, meaning that given a set of $N$ constants $g_n^{(1)}$, this system defines the set of $N$ constants $g_n^{(2)}$. Thus, it is possible to preserve the norm of all position vectors $\ket{e_k}$ by choosing the inner product appropriately.  Eq.~\eqref{eq:overlappres} yields a system of order $N^2$ equations with only $N$ constants to choose. Consequently, the overlap between the vectors $\ket{e_k}$ and $\ket{e_l}$ cannot in general be preserved between different representations and there seem to be properties of these vectors that inherently depend on the considered Hamiltonian. 
In other words, \emph{within this framework, the vector $\ket{e_n}$ cannot consistently represent the same physical state when the Hamiltonian is changed}; its physical meaning changes with the inner product. 

This means that Problem~\ref{prob:2} cannot be solved in its entirety by choosing a different inner product. We thus argue that the desired choice of inner product in non-Hermitian systems is the one presented in Eq.~\eqref{eq:finalgram}, providing a basis independent notion allowing for the mapping of representations of vectors between different Hilbert spaces, solving Problem~\ref{prob:1}.

\section{Discussion} \label{sec:discussion}

The problems with the biorthogonal formalism, listed in Sec.~\ref{sec:expvales}, indicate that conceptual and fundamental difficulties may arise when applying it in certain situations and setups; it is inconvenient in practice to work with an inner product that changes with the scaling of the eigenvectors of the Hamiltonian. As seen in Secs.~\ref{sec:indepIP}-\ref{sec:staterep}, Problem~\ref{prob:1} can be solved by describing the inner product using an inner product matrix. However, in contrast to Ref.~\cite{juNonHermitianHamiltoniansNogo2019}, we argue that it is important to specify the coefficients of the matrix $G$, denoted $g_n$ in Eq.~\eqref{eq:finalgram}, such that the inner product can be used in a simple way in physical setups. 
This provides a motivation for explicitly defining an inner product satisfying Constraints~\ref{const:1}-\ref{const:4}. Constraint~\ref{const:1} has its origins in physics, while Constraints~\ref{const:2}-\ref{const:4} are aimed at making the formulation straightforward to apply, and their combination hence allows us to define an inner product that is unique for a given Hamiltonians, without compromising its practicality.

We claim in Sec.~\ref{sec:indepIP} that the biorthogonal inner product can be recovered by choosing a particular case of the inner product given by Eq.~\eqref{eq:finalgram}. As the latter is expressed in terms of a matrix, and the former in terms of associated vectors, such a comparison is not obviously apparent. However, the inner product Eq.~\eqref{eq:finalgram} can be re-written in terms of associated vectors, facilitating such a comparison. Given a vector $\ket{\alpha} = \sum_n c_n(\sqrt{\braket{R_n}{R_n}})^{-1}\ket{R_n}$, the associated vector reads
\begin{equation}
    \ket{\tilde{\alpha}} = G\ket{\alpha} = \sum_n c_n\frac{\sqrt{\braket{R_n}{R_n}}}{\braket{R_n}{L_n}}\ket{L_n}.
\end{equation}
Importantly, and contrary to the associated vector in Eq.~\eqref{eq:associated_state}, this is independent of the choice of eigenvectors of $H$. 
The biorthogonal formalism is recovered when choosing $\braket{R_n}{R_n} = \braket{R_n}{L_n} = 1$. 
Using an inner product matrix has several advantages as it makes it easier to discuss several different inner products simultaneously. 
In addition, the matrix $G$ itself contains information about the system. 
For example, as stated in Sec.~\ref{sec:opalg}, we see that all observables, the self-adjoint operators in our formalism, have real eigenvalues.
In fact they correspond to operators that are pseudo-Hermitian with respect to $G$~\cite{delplaceSymmetryProtectedMultifoldExceptional2021}.

Addressing Problem~\ref{prob:2}, we showed in Sec.~\ref{sec:expvales} that this cannot be solved by modifying the inner product by any choice of $g_n$; in the biorthogonal setting the vectors $\ket{e_n}$ do not have meaning on their own without reference to an inner product. By extension the same holds true for all vectors; the vector space itself cannot describe physics without an inner product. 
We could leave it at that and say that it is pointless to discuss the physical meaning of the vectors $\ket{e_n}$, but since the biorthogonal bulk-boundary correspondence relies on information about gap closings provided by the expectation values of $\dyad{e_n}$, their physical significance cannot be disregarded. Furthermore, the experimental realization of the biorthogonal bulk-boundary correspondence is often done using classical systems, the mechanical system described in Ref.~\cite{ghatakObservationNonHermitianTopology2020} comprising a concrete example. 
In such systems the equations of motion can be rewritten as a Schrödinger equation of the form
\begin{equation}
    i\frac{d}{dt}\mathbf{x} = H\mathbf{x},
\end{equation}
where the vector $\mathbf{x}$ is not a position vector but instead contains both positions and velocities. 
Thus $\mathbf{x}$ contains physical quantities, while the matrix $H$ can be non-Hermitian and can therefore be thought of as a non-Hermitian Hamiltonian. 
In the particular case of Ref.~\cite{ghatakObservationNonHermitianTopology2020} the mechanical system maps to the non-Hermitian SSH-chain, for which a biorthogonal description implies that the vectors $\ket{e_n}$ do not have a physical meaning, cf., Eqs.~\eqref{eq:normpres} and \eqref{eq:overlappres}. 
However, since the vector $\mathbf{x}$ has physical meaning in the system, the vector $\ket{e_n}$ here clearly should have physical meaning of its own, without reference to an inner product. 
This implies that even though the biorthogonal expectation value can be used to find a bulk-boundary correspondence in these systems, it is uncertain whether such an inner product can be used for other things. There does not seem to be a one-to-one-correspondence between the physical quantities in biorthogonal quantum mechanics and the physical quantities in the mechanical systems mapped to non-Hermitian Hamiltonians. This suggests that biorthogonal quantum mechanics, even though it manages to predict gap closings in classical systems via the biorthogonal bulk-boundary correspondence, cannot be used to understand all the phenomena of these classical systems.

Even though the present study has mainly discussed discrete lattice models, it is apparent that similar interpretational problems can arise in the continuous case. 
In the context of wave functions, quantities such as $\braket{\mathbf{r}}{\psi}$ are usually referred to, which are non-trivial in a biorthogonal framework since the vectors $\ket{\mathbf{r}}$ suffer from the same interpretational problems as the vectors $\ket{e_k}$. That a biorthogonal inner product is not always suitable for describing the physics of a system is further supported by Ref.~\cite{silbersteinBerryConnectionInduced2020}, where it is explicitly stated that they prefer to use the expectation value used in Hermitian physics and not its biorthogonal counterpart.

\section{Conclusion and Outlook} \label{sec:conclusion}
In this work, we have studied the mathematical foundations of non-Hermitian quantum mechanics, focusing on problems arising from the definition of the biorthogonal inner product in the biorthogonal formalism. 
As pointed out in, e.g., Ref.~\cite{ibanezAdiabaticityConditionNonHermitian2014a}, the inner product leaves a scaling degree of freedom in how to choose the eigenvectors, a degree of freedom that we show affects physically relevant quantities including expectation values and transition probabilities. 
We have explained in some detail when this spare degree of freedom is important and when it is physically irrelevant. 
This clarifies why the biorthogonal formalism can be successfully applied in certain setups, e.g., when studying the biorthogonal bulk-boundary correspondence or when computing the Berry connection.

We have resolved this problem by defining a new, generalized, and basis independent inner product and by showing that one specific choice is favorable from a physical point of view.  We have explained how notions such as probability, observables and states should be modified, and shown that our formalism reduces to that of conventional quantum mechanics in the Hermitian limit. Furthermore, we have seen that the generalized inner product allows us to translate the meaning of states between different Hilbert spaces in a consistent manner.

We have discussed the position representation and physical consequences related to the vectors $\ket{e_n}$, whose physical meaning changes when the inner product changes. 
Unfortunately, this problem remains in the inner product formalism; simply changing the inner product does not allow for the physical meaning of vectors to be kept when mapped to different Hilbert spaces. 
We have concluded that a modified inner product can be used to \emph{translate} between different representations corresponding to the same physics, but that it is not sufficient to allow \emph{the same} vector as a representative of a physical state in different Hilbert spaces.
We expect similar problems for other representations that carry physical meaning.

A natural next step following this work would be to study the physical meaning of $\ket{e_n}$: 
Are there other methods that can be used in an attempt to retain a unified physical meaning of $\ket{e_n}$ in different Hilbert spaces, or is this generally impossible? 
The former would open up for a unification of the biorthogonal bulk-boundary correspondence and non-Hermitian quantum mechanics, while the latter indicates that the formalism itself is what implicitly assigns the physical meaning of $\ket{e_n}$, which on its own also would require further study.

The Hamiltonian operators considered in this work were assumed to be free from eigenvalue degeneracies. 
In particular, no exceptional points appear in their spectrum of the operators. 
Just as for the biorthogonal formalism, it is unclear how the formalism developed in this work extends to include defective operators. 
Given the attention of properties of exceptional points in the community, such an extension would comprise an additional way to study these exotic objects and increase the understanding of the fundamentals of non-Hermitian physics.

\emph{Note added}: After initial submission we were made aware of a line of research that addresses overlapping problems, see e.g. Ref.~\cite{mostafazadehConceptualAspectsPTSymmetry2010}.

\section*{Acknowledgements}
We thank Lukas R\o dland,  Emil J. Bergholtz and Jonas Larson for stimulating and fruitful discussions. We thank Eddy Ardonne for useful discussions and comments on the manuscript. E.E. and J.L.K.K. are supported by the Swedish Research Council (VR) and the Knut and Alice Wallenberg Foundation. J.L.K.K. is furthermore supported by the project Dynamic Quantum Matter (2019.0068) of the Knut and Alice Wallenberg Foundation.
\newpage
\bibliographystyle{iopart-num}
\bibliography{nhqm}{}

\end{document}